\documentstyle[aps,aipbook,floats,psfig]{revtex}
\def\etal{et al.}
\def\logp{log $N$-log $P$}
\def\logs{log $N$-log $S$}

\def\ith{{{i}${\underline {th}}$}~}

\def\Msol{M_\odot}
\begin{document}
\title{THE CORRECTED LOG $\bf N$-LOG FLUENCE DISTRIBUTION OF COSMOLOGICAL
$\bf \gamma$-RAY BURSTS}

\vspace{-0.45cm}
\author{Joshua S.~Bloom$^{1,2}$, Edward E.~Fenimore$^2$, Jean in 't
Zand$^{2,3}$} 
\address{
$^1$Harvard-Smithsonian Center for Astrophysics, Cambridge, MA  02138\\ 
$^2$Los Alamos National Laboratory, Los Alamos, NM 87544\\
$^3$Goddard Space Flight Center, Greenbelt, MD 20771}
\maketitle

\vspace{-0.37cm}
\begin{abstract}

Recent analysis of relativistically expanding shells of cosmological
$\gamma$-ray bursts 
standard and not peak luminosity ($L_0$).  Assuming a flat Friedmann
cosmology ($q_o = 1/2$, $\Lambda = 0$) and constant rate density
($\rho_0$) of bursting sources, we fit a standard candle energy to a
uniformly selected log~$N$-log~$S$ in the BATSE 3B catalog correcting
for fluence efficiency and averaging over 48 observed spectral shapes.
We find the data consistent with $E_0 = 7.3^{+0.7}_{-1.0} \times
10^{51}$ ergs and discuss implications of this energy for cosmological
models of $\gamma$-ray bursts.
\end{abstract}

\vspace{-0.4cm}
\section*{Introduction}
\vspace{-0.4cm}

On the basis of strong threshold effects of detectors, Klebesadel,
Fenimore, and Laros (1982) concluded that GRB fluence tests were
largely inconclusive.  As a result, nearly all subsequent
number-brightness tests have used peak flux ($P$) rather than fluence
($S$). However, the standard candle peak luminosity assumption that is
required by \logp\ studies is unphysical.  If, for instance, bursts
originate at cosmological distances and are produced by colliding
neutron stars\cite{Pac90} then one might expect that total energy
would be standard and not peak luminosity.  Moreover, recent analysis
of the time histories in relativistically expanding shell models has
found the required differences in bulk $\Gamma$ factor between
different GRBs all but eliminates the possibility of a standard candle
luminosity in such models\cite{Mad96}.

In this paper, we seek to eliminate the large threshold effects
present in \logs\ studies by correcting the observed number of bursts
at a given fluence by the trigger efficiency of the detector.  In \S I
we use the calculated trigger efficiency in PVO\cite{Jean96} and the
catalogue of PVO events\cite{PVO96} to test the correction
algorithm. In \S II we examine the fitting algorithm to the \logs\
curve in BATSE 3b and find a standard candle energy for cosmological
gamma-ray bursts.  In \S III we discuss the implications of such an
energy and the distances implied by the fit.

\vspace{0.2cm}
\centerline{\bf I.~PVO CONSISTENCY CHECK}
\vspace{0.2cm}

The Pioneer Venus Orbiter (PVO) had a peak flux trigger system
(sampled on 0.25, 1.0, and 4.0 sec timescales) and was sensitive to
bursts down to fluxes of $5\times10^{-6}$ erg cm$^{-2}$.  Despite a
substantially lower fluence trigger sensitivity range, PVO saw
hundreds more bright bursts than BATSE due the relatively long on-time
and large sky-coverage of PVO.  As the bright region of the BATSE
\logs\ curve seems to fit a -3/2 power law well, we would expect that
the entire \logs\ curve of PVO should show a similar behavior.

Using the PVO trigger efficiency, $\epsilon(S)$, from in t' Zand \&
Fenimore (1996), for each burst $i$ with fluence $S_i$ in the PVO catalogue
we take the expected number of bursts with pre-detection fluence to be
$N_{i,\rm true}= N_{i, \rm{obs}}/\epsilon(S_i)$.  We then compare the
derived log $N(>S)$-log $S$ curve with a -3/2 power law with arbitrary
$S$-intercept as seen in figure~(1b).  Although at lower fluence there
appears to be a deviation from -3/2, the fit is good: with a
Kolmogorov-Smirnov (KS) probability of 40\% that the corrected
distribution comes from a -3/2 power law.  We derive this KS
statistic by finding the maximum distance between the corrected and
-3/2 distributions (in linear space) down to $S=10^{-4.5}$ erg
cm$^{-2}$, the fluence at which $\epsilon(S)$ falls to 50\%.  By
comparing the distributions down to lower fluences, we find an even
better KS fit. This is understandable because as we add more bursts to
the distribution, small deviations at high fluences contribute less to
the KS distance parameter. Note that although the fit is acceptable, a -3/2
slope is not necessarily required by the corrected data.  
We thus conclude that the trigger efficiency
determination algorithm from in t' Zand and Fenimore (1996) is sound,
at least in PVO.

\vspace{0.2cm}
\centerline{\bf II.~Deriving the log $\bf{N}$-log $\bf{S}$ Curve}
\vspace{-0.5cm}
\subsection{BATSE Trigger Efficiency}
\vspace{-0.5cm}
One subtly worth noting is that BATSE trigger efficiencies are model
dependent, ie.~they depend on the choice of $E_0$ and cosmology, since it is
necessary to know {\it a priori} the true underlying distribution of
bursts that passes by the detector.

\begin{minipage}{2.5in}
  \psfig{file=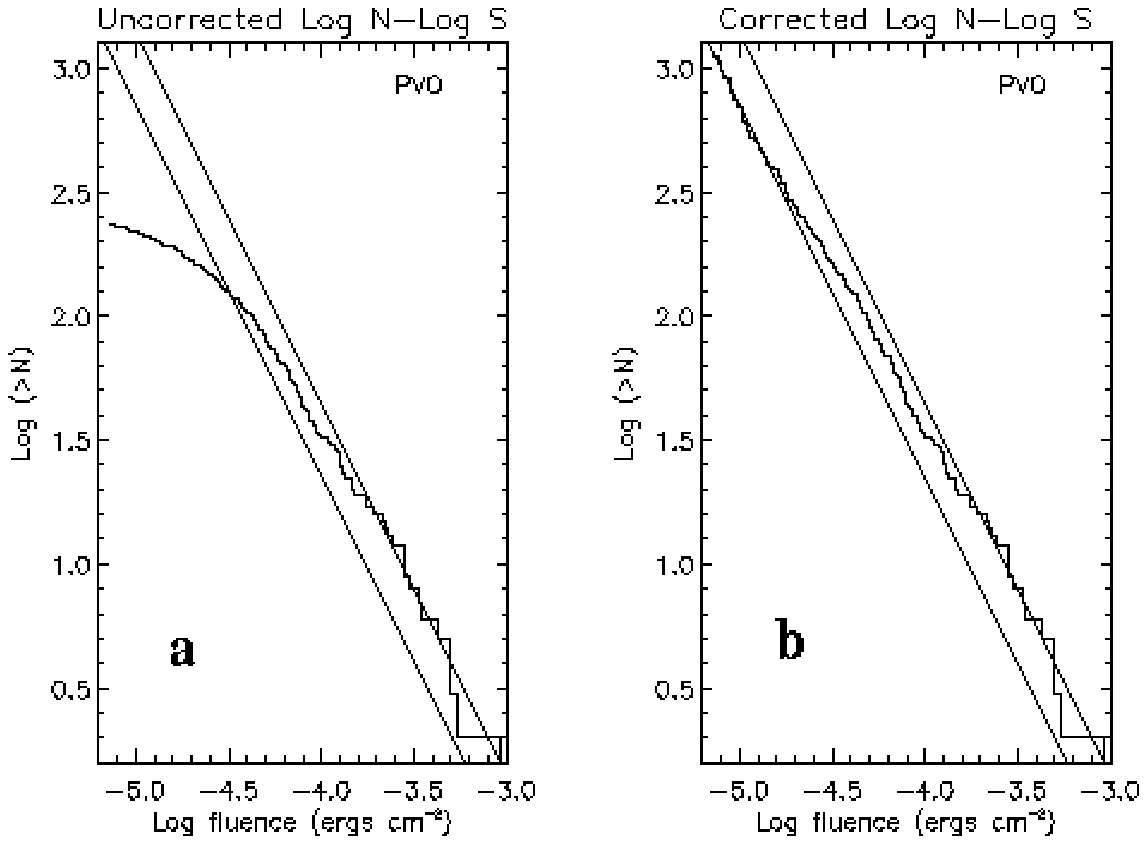,height=3.0in,width=2.5in,angle=0}
\end{minipage} 
\begin{minipage}{1.9in}
{\bf FIG.~1} {The log $N$-log $S$ curve for PVO. a$)$ The uncorrected
curve for 293 events in the PVO catalogue which shows significant
departure (KS $\simeq 0.0$) from the -3/2 power law (shown as solid
line) expected from BATSE observations. b$)$ The corrected curve (KS =
0.40 at $\epsilon(S) = 0.50$) using the PVO trigger efficiency from in
t' Zand \& Fenimore (1996).  Any deviation from a -3/2 power-law at
low fluences we attribute to an incomplete understanding of the
trigger efficiency.}
\end{minipage}

\noindent In fact, the
derivation of the PVO $\epsilon(S)$ assumed an underlying -3/2
distribution.  Petrosian and Lee (1996) have constructed trigger
efficiencies using bivariate correlation.  While this method does not
assume a particular cosmology, it does require that GRB brightness and
duration are inherently uncoupled.  Our method does not have this
requirement and we make no assumptions about the bursts other then
they are cosmological in origin.  The BATSE trigger efficiencies could
be calculated for any $E_0$ ($q_0 = 1/2$, $\Lambda = 0$) and two are
depicted in figure (2a). Note that the efficiency is nearly unity for
the several orders of magnitude in fluence.  The corrected \logs\
curve for BATSE is depicted in figure (2b) for two values
of $E_0$. Interestingly, the two distributions are nearly identical for
most of the fluence range.  In addition, it is clear
that the bend from -3/2 in \logs\ is true.

\vspace{0.2cm}
\centerline{\bf B.~Standard Candle Energy Fits}
\vspace{0.2cm}

The observed fluence of a source depends strongly on the spectrum, and
since the observed spectral shape depends on the distance to the
object, the intrinsic spectrum of a GRB object must be used.  In
addition, the normalization and the spectral shape vary over the
duration of the burst, adding to the uncertainty in analysis.  

Following a similar analysis as in Fenimore and Bloom (1995), we take
as our baseline spectra averages over the GRB spectra fit by Band
\etal\ (1993).  Each such baseline burst has associated with it an observed
fluence, $S_{i}$, and an observed spectral shape, $\phi_i(E)$.  

\begin{minipage}{4.5in}
\centerline{\vspace{-5.0cm}a\vspace{5.0cm}\hbox{
    \psfig{file=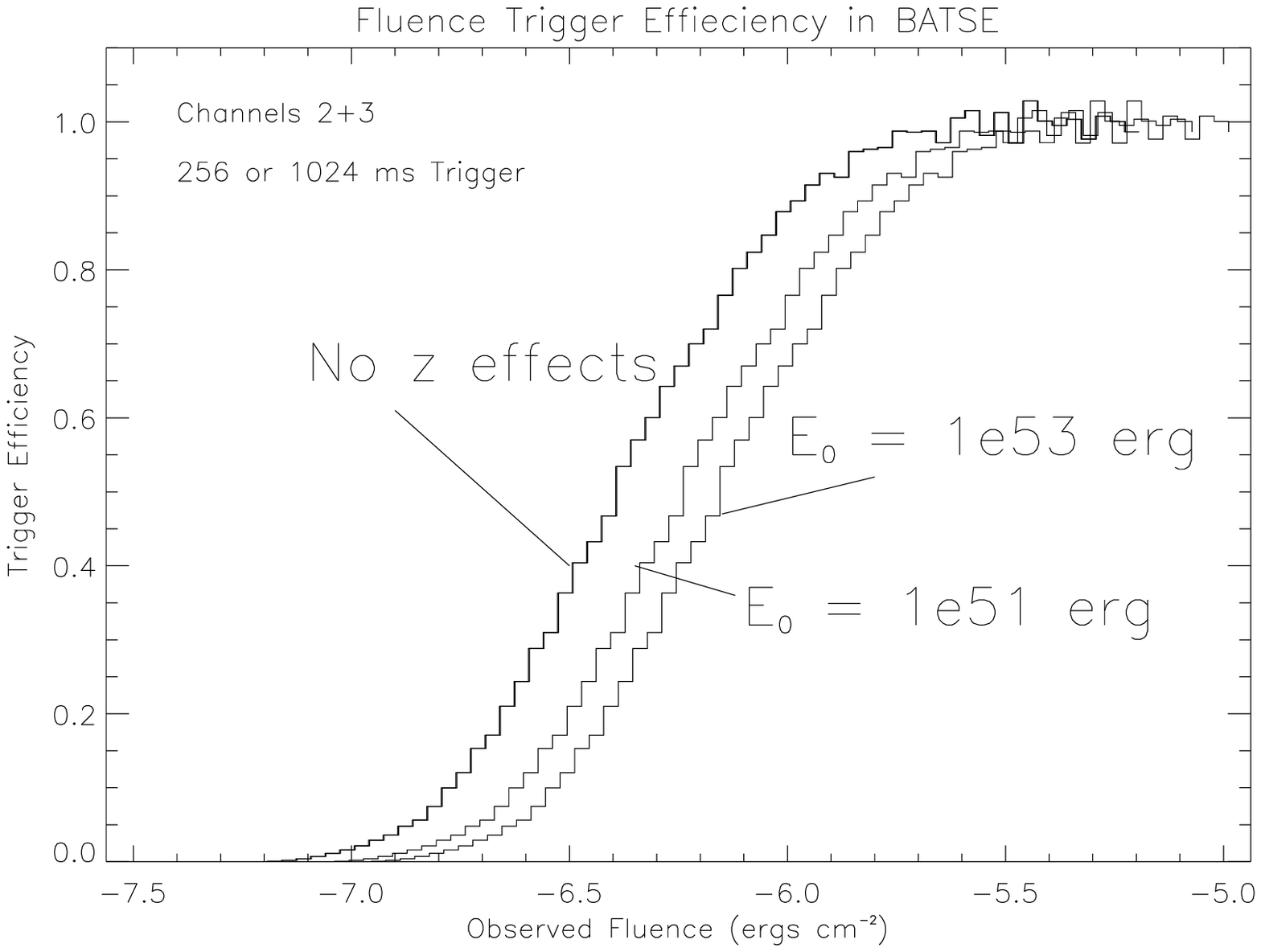,height=2.5in,width=2.45in,angle=0} 
    \psfig{file=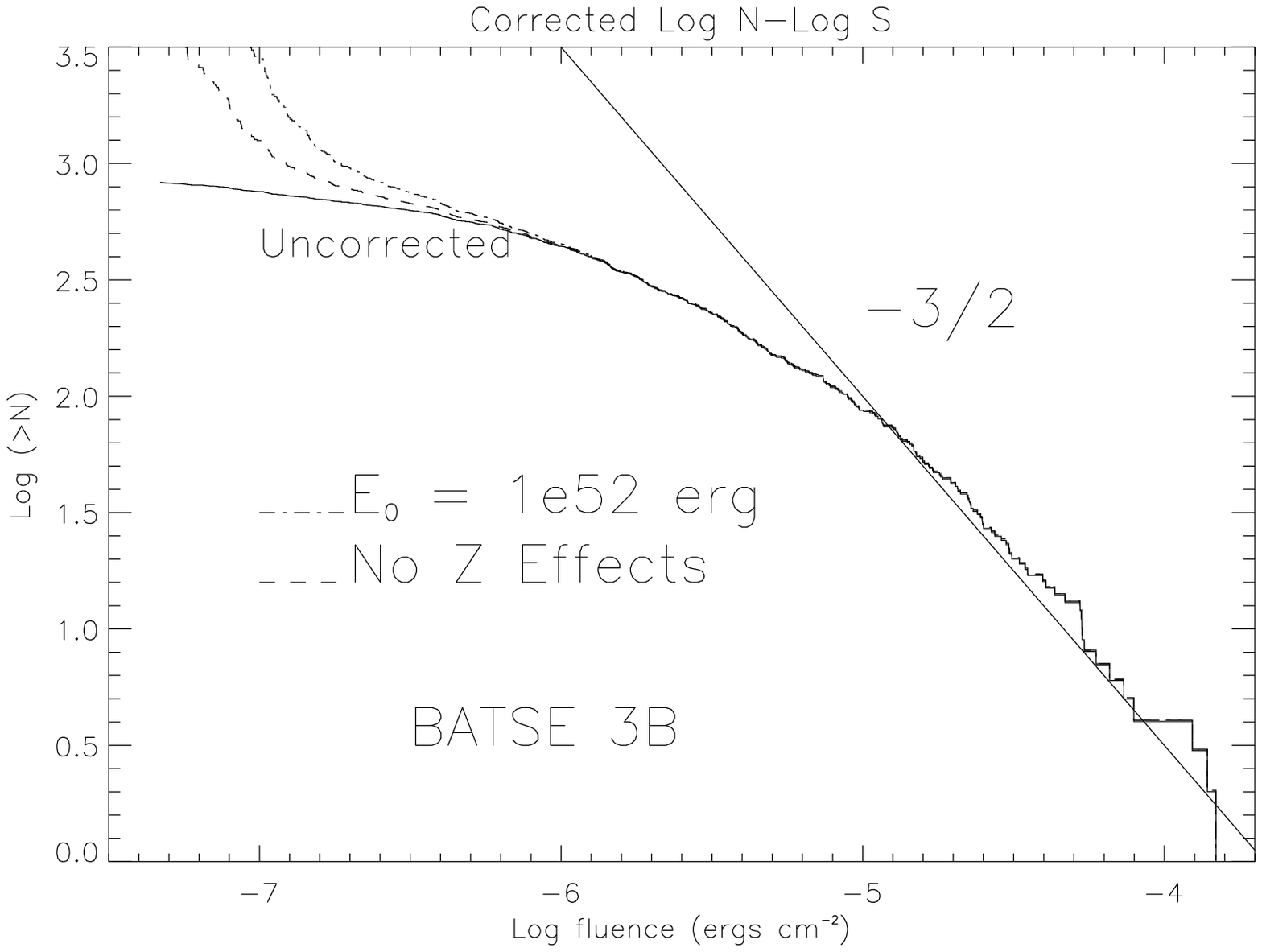,height=2.5in,width=2.45in,angle=0} 
}b}
\begin{minipage}{4.6in}
{\bf FIG.~2:} {a) The BATSE trigger efficiency for different $E_0$
and b)  the corrected \logs\ curve for 830 BATSE bursts.  Along with
the uncorrected \logs\ curve, we depict a corrected curve
corresponding to an assume standard candle energy of $E_0$ of
$10^{52}$ ergs (dot-dash line) and a corrected curve where the effect
of redshift on the baseline spectra is removed (dash line). }
\end{minipage}
\end{minipage}

\vspace{0.3cm}

\noindent 
Since
each Band \etal\ (1993) burst spectrum is averaged over the duration
of the burst, we assume in the following analysis that the
spectral shape is constant, that is, $\phi(E,t_s) \simeq N(t_s)
\phi_i(E)$.  The fluences, $S_{i}$ [ergs cm$^{-2}$], are
available for 48 of the Band \etal\ (1993) bursts in BATSE 3B\cite{Meegan96}. 

The observed spectral shape, $\phi_i(E)$, will not necessarily come
from a burst at $z \sim 0$ especially if $E_0$ is large.  Therefore,
for a given $E_0$, $S_{i}$, and $\phi_i(E)$ we first solve for the
redshifts, $z_{i}$, of the baseline events associated with each spectral shape.
The standard candle energy, $E_0$, is given by,
\begin{equation} E_0 = 4\pi R^2_{i,z} \int_0^\infty N(t_s) dt_s 
	\int_{30}^{2000} E \phi_i\left(E \over {1 + z_{i}}\right) dE  
	\label{SC_E}\end{equation}
where $N(t_s)$ is the normalization of the spectrum (units of ergs
keV$^{-1}$) as a function of time at the source.  The comoving
distance,  $R_{i,z}$, is defined in eq.~[2] of Fenimore \& Bloom
(1995). The energy range used in calculating $E_0$ in eq.~(\ref{SC_E}) is
taken as 30-2000 keV, since we later compare $E_0$ to
standard candle peak luminosity found in the same energy band.  

The observed fluence of the \ith baseline burst in the energy
range 50-300 keV is given by,

\vspace{-0.35cm} 

\begin{equation}S_{i} = \int_0^\infty N(t_{obs}) dt_{obs} 
	\int_{50}^{300} E \phi_i\left[\frac{1+z_r}{1+z_i}E\right] dE, 
\label{def_fluence}\end{equation}
\noindent where $N(t_{obs})$ is the observed normalization of the spectrum.  

For a given standard candle energy, $E_0$, we numerically determine
the redshift ($1+z_{i}$) of the \ith baseline burst using
eqs.~(\ref{SC_E}, \ref{def_fluence}) and letting $z_r = z_i$.
Note that $(1+z_i)\int N(t_s) dt_s = \int N(t_{obs}) dt_{obs}$.

Instead of assuming a spectral shape at the source, we use an average
over baseline spectra to compute the number of expected observed
bursts, $\Delta N_{exp}[S_j {\rm~to~} S_{j+1}]$ in some fluence range
[$S_j, S_{j+1}$]:
\begin{equation}
\Delta N_{exp}[S_j {\rm~to~} S_{j+1}] = \frac{4\pi}{N_{BAND}} \sum_{i=1}^{N_{BAND}} 
\int_{R(S_j)}^{R(S_{j+1})} \epsilon[S_i(r)] 
\frac{\rho_0}{1+z_r} r^2 dr.  \label{NUMbin}
\end{equation}
where $N_{BAND} = 48$ is the number of baseline spectra used and
$\rho_o$ is the rate density of bursts per comoving volume.  The
quantity $S_i(r)$ is the predicted fluence (using eqs.~[\ref{SC_E},
\ref{def_fluence}]) of the \ith baseline burst if it was at a distance
$r$.  This distance corresponds to a redshift $1+z_r$.

We construct 11 fluence bins (in BATSE channels 2+3 corresponding to
approximately 50-300 keV) of roughly equal number of bursts.  We
select bursts with $C_{min}/C_{max} > 1$ on either the 256 or 1024 ms
timescale, then find a minimized $\chi^2$ between the number of
predicted bursts and observed by varying $E_0$.  For 9 degrees of
freedom we find an acceptable $\chi^2 = 14.7$ corresponding to a
standard candle $E_0 = 7.3_{-1.0}^{+0.7} \times 10^{51}$ ergs.  Table
(\ref{goodfit}) gives the bin ranges, number of observed bursts per
bin, number of predicted bursts for the best fit energy, and their
implied redshifts.

\begin{table}[tb]
\vspace{-0.8cm}
\caption{Best Fit Distribution of  $E_0 = 7.0 \times 10^{51}$ ergs}
\label{goodfit}
\begin{tabular}{cccccc}
&\multicolumn{2}{c}{Fluence Ranges$^a$(50-300 keV)}&
\multicolumn{2}{c}{$\Delta N[S_j {\rm~to~} S_{j+1}]$} \\
Bin Number($j$) & $S_j$ & $S_{j+1}$ 
& Observed$^b$ & Predicted & $1+z_j$ \\
\tableline
1  & 2.16e-07  & 3.82e-07   & 51 & 38.4 & 3.88\\
2  & 3.82e-07  & 5.85e-07   & 42 & 50.5 & 3.24\\
3  & 5.85e-07  & 7.55e-07   & 42 & 36.3 & 2.84\\
4  & 7.55e-07  & 1.13e-06   & 46 & 63.0 & 2.64\\
5  & 1.13e-06  & 1.43e-06   & 37 & 36.8 & 2.36\\
6  & 1.43e-06  & 2.00e-06   & 48 & 49.7 & 2.22\\
7  & 2.00e-06  & 2.80e-06   & 39 & 44.3 & 2.04\\
8  & 2.80e-06  & 4.05e-06   & 44 & 41.1 & 1.89\\
9  & 4.05e-06  & 6.20e-06   & 37 & 37.2 & 1.74\\
10 & 6.20e-06  & 1.36e-05   & 40 & 44.7 & 1.60\\
11 & 1.36e-05  & 6.60e-05   & 41 & 32.3 & 1.41\\
\end{tabular}
\noindent{$^a$ In ergs cm$^{2}$}

\noindent{$^b$ Bursts with $C_{min}/C_{max} > 1$ on the 256 or 1024 ms
timescale in BATSE 3b.}
\vskip -7pt
\end{table}

\vspace{0.2cm}
\centerline{\bf III.~CONCLUSIONS}
\vspace{0.2cm}

Our fit of $E_0 = 7.0_{-1.0}^{+0.7} \times 10^{51}$ [30-2000 keV] ergs
seems a plausible number on the basis that GRBs last on the average 10
sec and $L_0 = 4.6 \times 10^{50}$ erg s$^{-1}$ from \logp\
studies\cite{Bloom95}.  However, this $E_0$ implies a rather large
efficiency of energy conversion to $\gamma$-rays ($\sim 10\% $) if the
bursting mechanism is colliding neutron stars ($M_{total} \simeq 2.8
\Msol$).  Nevertheless, this result would seem to help resolve the
``no-host'' problem (cf.~Fenimore 1993 \etal).  Interestingly, that
the dimmest bursts ($S \simeq 5 \times 10^{-8}$ erg cm$^{-2}$) are
required to be at a redshift of $1 + z \simeq 6.4$ given this $E_0$,
would seem to rule out several cosmological models that require GRB
progenitors to be within galaxies (although see Lu \etal\ 1996).  This
surprisingly high redshift is due to the correct blueshifting of the
baseline spectra back to the source in eq.~(1).  If we neglect this
factor, we obtain a smaller, more tenable redshift of the dimmest
bursts ($1+z = 5.2$).

Whatever the conclusion about the models, we note two important
results.  First, the bend in the \logs\ curve in BATSE is real, not
an artifact of strong threshold effects.
This implies that we are seeing
either a truncated spatial distribution of GRBs (as in Galactic models) or an effect due to the expansion of the universe.
The bend might also be caused by a combination of rate density or
number density evolution, and a study of their possible effects is
certainly warranted.  Secondly, with the availability of Monte Carlo
modeling of trigger efficiencies, \logs\ tests need no longer be
inconclusive.

\end{document}